%% file: cover.tex
\RequirePackage{rotating}
\documentclass[format=acmsmall, review=false, screen=true]{acmart}
\renewcommand\footnotetextcopyrightpermission[1]{} % removes footnote with conference information in first column
\usepackage{booktabs} % For formal tables
\usepackage{rotating} % Rotating table
\usepackage[ruled]{algorithm2e} % For algorithms
\usepackage{multirow} % For multi-rows

\SetAlFnt{\small}
\SetAlCapFnt{\small}
\SetAlCapNameFnt{\small}
\SetAlCapHSkip{0pt}
\IncMargin{-\parindent}

\fancypagestyle{firstpagestyle}{\fancyhf{}} %Remove Footer and Page label in Cover Page

\usepackage{tabularx}

\usepackage{xcolor}

\usepackage{url}

\begin{document}
% Title portion. Note the short title for running heads 
\title{Opportunities and challenges for deep learning in cell dynamics research}  

\author{Binghao Chai}
\affiliation{%
  \institution{Queen Mary University of London, United Kingdom}
}

\author{Christoforos Efstathiou}
\affiliation{%
  \institution{Queen Mary University of London, United Kingdom}
}

\author{Haoran Yue}
\affiliation{%
  \institution{Queen Mary University of London; Currently, University of Sussex, United Kingdom}
}

\author{Viji M. Draviam}
\affiliation{%
  \institution{Queen Mary University of London and The Alan Turing Institute, London, United Kingdom}
}

\begin{abstract}
ABSTRACT 
\\
\\ With the growth of artificial intelligence (AI), there has been an increase in the adoption of computer vision and deep learning (DL) techniques for the evaluation of microscopy images and movies. This adoption has not only addressed hurdles in quantitative analysis of dynamic cell biological processes, but it has also started supporting advances in drug development, precision medicine and genome-phenome mapping. Here we survey existing AI-based techniques and tools, and open-source datasets, with a specific focus on the computational tasks of segmentation, classification, and tracking of cellular and subcellular structures and dynamics. We summarise long-standing challenges in microscopy video analysis from the computational perspective and review emerging research frontiers and innovative applications for deep learning-guided automation for cell dynamics research.
\end{abstract}

\maketitle

\input{body}

\clearpage
\input{glossary}

% Bibliography
\newpage
\bibliographystyle{unsrt}
\bibliography{bibliography}

\end{document}

%% file: body.tex
% Project Context and Background
\section{Introduction}
Advances in microscopy have influenced a range of cell biology and biomedical research areas. Microscopy advances supported by automated or semi-automated image analysis are being transformed by deep learning (DL) approaches. DL approaches for the analysis and restoration of microscopy image datasets have been reviewed recently~\cite{xu2022deep,moen2019deep,krentzel2023deep,choi2021emerging,hallou2021deep,hollandi2022nucleus}, but there is no comprehensive survey of the status of artificial intelligence (AI) methods for tracking or predicting trajectories of dynamic structures in microscopy movies. Time-lapse movies of dynamic cell biological processes are particularly a unique case because of the temporal discontinuity in image acquisition which is being offset through high-speed and volumetric imaging~\cite{efstathiou2021electrically,liu2023characterization,mimori2021imaging}. Machine learning and deep learning (ML/DL) methodologies that demonstrate superior performance in most image analysis tasks need to be adapted for movie analysis tasks. 

Implementing DL approaches involves data annotation, denoising, selection and training of a chosen neural network, evaluating and optimising the DL model, and assessment of outcomes - all dependent on specific imaging and analysis tasks. For a practical guide on how to build DL models for image analysis, we direct readers to reviews focussing on bioimage analysis workflows~\cite{gomez2022building,janowczyk2016deep,lv2022deep}.
 
In this review, we present an in-depth survey of current AI-based microscopy image and movie analysis considering three key computational tasks: object segmentation, classification and tracking. We contrast conventional image analysis approaches against DL techniques (neural network architectures) that have been successfully used in cell biology. To benefit future DL tool development, we collate a list of existing open-sourced datasets. Throughout we discuss accurate and efficient ways of data preparation for use in DL applications. Finally, we highlight critical challenges and limitations with current deep learning applications in analysing dynamic cell biology movies, along with possible opportunities for future research. 

\section{AI-guided advances in image analysis}
We open with a list of successes in microscopy image analysis brought in through machine learning or deep learning (ML/DL) methods and list how these can set new trends in cell biology:

\begin{itemize}
\item Analysing large image datasets in a context-free and efficient way: Ideal for large time-lapse videos or genome-wide imaging screens.
\item Automation of computational tasks: Image segmentation, classification, tracking and transformation support high-fidelity spatiotemporal studies of cellular processes.
\item Learn/recognise complex structures: Recovering hidden patterns amidst known morphological features for hypothesis-building and better data interpretation.
\item Managing noise and variation: Handling morphological and intensity variations, can bolster data reproducibility and reduce the chances of human biases or errors.
\end{itemize}

{
\footnotesize
\begin{table}[!htbp]
\centering
\caption{\label{tab:dl_tech}Deep learning techniques for cell biology.}
\begin{tabular}{p{4.5cm}p{8cm}}
\toprule
\bfseries{Deep learning techniques} & \bfseries{Application in microscopy data analysis} \\
\midrule
Convolutional Neural Networks~\cite{lecun2015deep,krizhevsky2017imagenet} & Segmentation~\cite{raza2019micro}, classification~\cite{khan2019novel,coudray2018classification,alzubaidi2020deep,shahin2019white,jha2019mutual}, tracking~\cite{newby2018convolutional,lugagne2020delta} \\
Recurrent Neural Networks~\cite{sherstinsky2020fundamentals} & Cell tracking~\cite{newby2018convolutional}, segmentation~\cite{wollmann2019gruu}, cell cycle analysis~\cite{jose2023automatic,kimmel2019deep} \\
U-net~\cite{ronneberger2015u} & Segmentation~\cite{karabaug2023impact,joseph2020quantitative,lugagne2020delta,gomez2019deep,lugagne2020delta,ronneberger2015u,cciccek20163d,falk2019u}, denoising~\cite{ahmed2021medical}, feature extraction~\cite{yang2021multi} \\
Generative Adversarial Networks \cite{goodfellow2020generative} & Image denoising~\cite{fuentes2022mid3a,zhong2021blind,chen2021three}, data augmentation~\cite{naghizadeh2021semantic,majurski2019cell,yu2021generative,chlap2021review}, virtual staining of biological samples~\cite{bai2023deep} \\
Graph Neural Networks~\cite{wu2020comprehensive} & Cell tracking~\cite{moen2019deep} \\
\bottomrule
\end{tabular}
\end{table}
}

% A variety of DL techniques and tools have been successfully implemented for microscopy image analysis (Table~\ref{tab:dl_tech}). 
% The most widely used DL method for object  classification~\cite{khan2019novel,coudray2018classification,alzubaidi2020deep,shahin2019white,jha2019mutual}, segmentation~\cite{raza2019micro} and tracking~\cite{newby2018convolutional,lugagne2020delta} is the Convolutional neural network (CNN)~\cite{lecun2015deep,krizhevsky2017imagenet}. U-net was specifically designed for image segmentation and is now a widely used model successfully applied to biomedical image segmentation and object detection~\cite{ronneberger2015u}. Recurrent neural network (RNN) is another type of deep learning model that can process sequences of data including time-lapse images of microscopy movies. RNNs have been used for tasks such as cell tracking~\cite{sherstinsky2020fundamentals}, event detection and cell cycle analysis~\cite{jose2023automatic,kimmel2019deep}. Generative adversarial network (GANs)~\cite{goodfellow2020generative} is a deep learning technique which can generate new data that is similar to a given set of input data. GAN may be used for data augmentation to solve the data shortage problem in many scenarios for instance, due to 3D volumetric stacks, photobleaching or phototoxicity~\cite{fuentes2022mid3a,zhong2021blind,chen2021three,naghizadeh2021semantic}. 

\begin{figure}[!htbp]
\centering
\includegraphics[width=11.2cm]{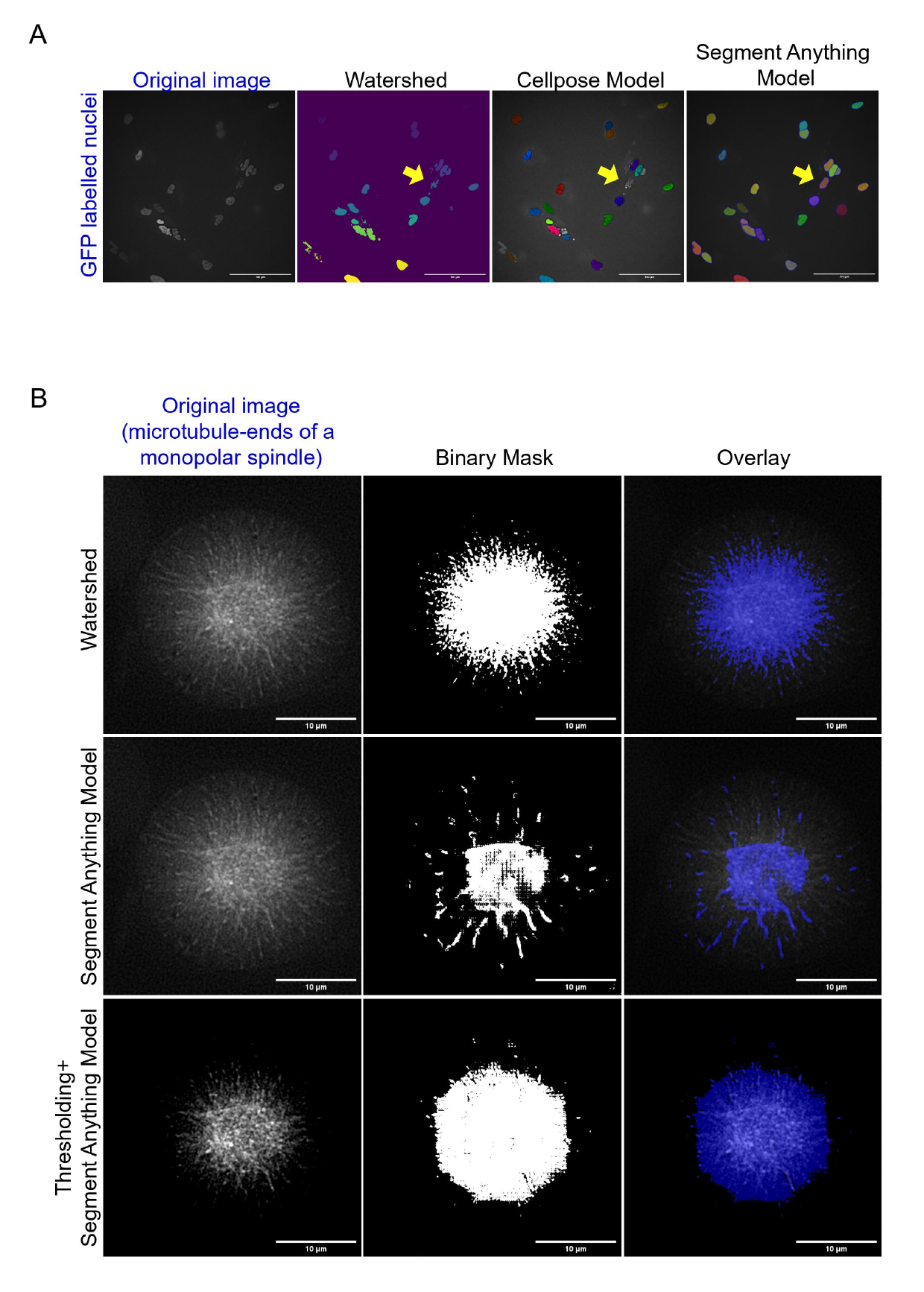}
\caption{Image segmentation outcomes with zero-cost training for distinct subcellular objects show different outcomes. (A) shows the original unprocessed image of RPE1 cells with p21-GFP labelled nuclei displaying a variety of nuclear shapes (normal and abnormal), and segmentation outcomes using conventional watershed segmentation,  CellPose and SAM, as indicated.   Successfully segmented nuclei are pseudocoloured in different colours. Yellow arrows indicate rare nuclear instances that are differently segmented using the three approaches.
(B) shows the original image of a single monopolar spindle decorated by EB3-mKate2 comets at microtubule-ends (unprocessed or with intensity thresholding), segmented binary masks using SAM or Watershed, and their overlay on respective raw images, as indicated. SAM and watershed segmentation differently segment EB3-mKate2 comets.  In the top row, the original image was normalized to a range between 0 and 1. Thresholding was aimed at separating astral microtubules from background cytoplasmic noise.}
\label{fig:new_fig}
\end{figure}

Table~\ref{tab:dl_tech} displays the most widely used DL techniques for microscopy image analysis. Apart from these well-established techniques, a reusable and adaptable image segmentation architecture, the Segment Anything Model (SAM), has been proposed by Meta AI which is a zero-shot transfer learning approach~\cite{kirillov2023segment}. Its performance appears to be competitive with or even superior to prior fully supervised trained models and is applied in medical imaging~\cite{he2023accuracy,mazurowski2023segment} and digital pathology~\cite{deng2023segment}. SAM is unexplored for cellular or subcellular segmentation tasks. We tested SAM's initial features on our data. Fig~\ref{fig:new_fig} (panel A) showcases SAM's capability to segment complex nuclear morphologies without additional training. Some under/over-segmentations occur, but they are comparable to watershed segmentation (conventional) and CellPose (DL), making SAM a tool that could save considerable researcher time. However, it encounters challenges with intricate subcellular structures~\cite{wang2023empirical}. For example in the case of microtubule-end segmentation, while the conventional watershed method can partially segment microtubule-ends found in a monopolar spindle, SAM segments the whole monopolar spindle instead (Fig~\ref{fig:new_fig}.B). Evidently, SAM holds the capability to simplify segmentation, but it has not yet been tested in densely packed microscopy images. For instance, electron microscopy (EM) images displaying crowded organelles may pose challenges to achieving accurate segmentation without trained datasets of individual organelles.

\section{AI-guided methods outperform conventional image analysis tools}
DL neural networks are more effective than traditional computer vision techniques. They learn from large-scale datasets and have the capacity to extract high-level features without heavy reliance on domain knowledge for feature extraction~\cite{o2020deep}. While many DL tools have focused on segmenting nuclei and whole cells with fluorescently-labelled markers, some specialised DL tools have been developed to segment distinct organelles such as Golgi apparatus, mitochondria and endoplasmic reticulum from Electron Microscopy (EM) data (Table~\ref{tab:subcellular}). However, DL tools that can both segment and track dynamic subcellular structures in time-lapse fluorescent movies are currently limited. Mitochondria~\cite{lefebvre2021automated}, microtubule-ends and mitotic spindles~\cite{dang2023deep} are among the few dynamically changing structures for which automated analysis tools are available, but deep learning has only been used in the last case. The most popular DL-based tools include U-net~\cite{ronneberger2015u,cciccek20163d,falk2019u}, StarDist~\cite{schmidt2018cell,weigert2020star} and Cellpose~\cite{stringer2021cellpose,pachitariu2022cellpose,schiff2022integrating}. Most DL-based solutions are data-driven and so there are no standards to inform biologists which model is the most suitable one for their own dataset for specific computational tasks. As a result, most people veer towards integrated platforms such as Fiji (through plugins)~\cite{schindelin2012fiji,gomez2021deepimagej}, CellProfiler~\cite{mcquin2018cellprofiler}, QuPath~\cite{bankhead2017qupath}, ZEISS arivis Cloud (formerly APEER)~\cite{apeer2023,david2021apeer} and ZeroCostDL4Mic~\cite{von2021democratising, zerocostdl4micimplementation}. Below we discuss the application of deep learning in cellular image and movie analysis: segmentation, classification, and tracking, and contrast it against conventional non-DL methods.

\subsection{Segmentation}
Two types of image segmentation, semantic and instance, serve different purposes. Semantic segmentation aims to classify individual pixels within an image into specific classes. It groups instances of a class together, lacking the ability to differentiate individual instances like overlapping nuclei. However, it effectively separates membrane outlines from intra or extracellular space. Instance segmentation differentiates instances of the same class (In Fig~\ref{fig:new_fig}.A, Cellpose~\cite{stringer2021cellpose,pachitariu2022cellpose} and Segment Anything Model (SAM)~\cite{kirillov2023segment} separate overlapping nuclear objects could be separated as distinct instances). Recently, panoptic segmentation has been introduced which combines instance and semantic segmentation where each instance of an object in the image is segregated and the identification of each object is predicted~\cite{liu2020unsupervised}.

Conventional segmentation methods include thresholding, edge-based algorithms and region-based segmentation~\citep{minaee2021image}. Edge-based segmentation like \textit{Canny} or \textit{Sobel} edge-detectors followed by contour filling~\cite{minaee2021image}performs better than thresholding, but can produce imperfect contours. Region-based segmentation, particularly watershed segmentation is widely used in cell biology~\cite{minaee2021image}. Conventional segmentation methods are often used for automated annotation of large datasets, followed by manual correction to save annotation time~\cite{dang2023deep}.

\begin{sidewaystable}[!htbp]
\centering
\scriptsize
\caption{\label{tab:subcellular}Deep Learning-based (DL-based) tools available for subcellular organelle segmentation. Tools with comprehensive documentation and tutorials that can be accessed, independent of the source code, are highlighted in \textcolor{green}{green}.}
%\begin{tabular}{p{1.5cm}p{3.5cm}p{2cm}p{5.5cm}p{1.5cm}p{3cm}p{2.5cm}}
\begin{tabular}{p{2cm}p{4.5cm}p{3cm}p{1.5cm}p{5cm}p{2.5cm}}
\toprule
\bfseries{Tool} & \bfseries{Subcellular Structures} & \bfseries{DL Architecture} & \bfseries{Dynamics Tracking} & \bfseries{Strengths in User Experience} & \bfseries{Source} \\
\midrule
\textcolor{green}{U-Net}~\cite{ronneberger2015u,cciccek20163d,falk2019u} &
  Fluorescent and label-free cell membrane, fluorescent nuclei and EM neurites &
  CNN &
%  Object segmentation. &
  No &
  Documentation (application), tutorials (Jupyter notebook). &
  \url{https://lmb.informatik.uni-freiburg.de/people/ronneber/u-net/} \\
\textcolor{green}{Cellpose}~\cite{stringer2021cellpose,pachitariu2022cellpose} &
  Fluorescent cell membrane and nuclei &
  U-net &
% A generalist algorithm for nucleus and cell membrane segmentation. &
  No &
  Documentation (installation and application), tutorials (Jupyter notebook), integrated through the ZEISS arivis Cloud (formerly APEER platform)~\cite{apeer2023,david2021apeer}. &
  \url{https://github.com/mouseland/cellpose} \\
\textcolor{green}{Stardist (3D)}~\cite{schmidt2018cell,weigert2020star} &
  Fluorescent and H\&E stained nuclei &
  U-net &
%  Object detection. &
  No &
Documentation (installation and application), tutorials (Jupyter notebook), integrated plugin (ImageJ/Fiji~\cite{schindelin2012fiji}, Napari~\cite{napari}, QuPath~\cite{bankhead2017qupath}). &
  \url{https://github.com/stardist/stardist} \\
ASEM~\cite{gallusser2022deep} &
  EM Golgi apparatus, mitochondria, nuclear pore complexes, caveolae, endoplasmic reticulum, clathrin-coated pits, vesicles &
  3D U-net &
%  Automated segmentation of subcellular structures in Electron Microscopy. &
  No &
  Documentation (installation and application). &\url{https://github.com/kirchhausenlab/incasem} \\
\textcolor{green}{MitoSegNet}~\cite{fischer2020mitosegnet} &
  Fluorescent mitochondria &
  U-net &
%  Segmentation of mitochondria in 2D. &
  No &
  Documentation (application), GUI (multiple operation systems). &
  \url{https://github.com/MitoSegNet/MitoS-segmentation-tool} \\
\textcolor{green}{SpinX}~\cite{dang2023deep} &
  Fluorescent mitotic spindle, label-free cell cortex &
  Mask R-CNN &
%  Track the dynamics of subcellular structures in 3D time-resolved movies. &
  Yes &
  Documentation (application), integrated through the ZEISS arivis Cloud (formerly APEER platform)~\cite{apeer2023,david2021apeer}. &
  \url{https://github.com/Draviam-lab/spinx\_local} \\
Multicut~\cite{beier2017multicut} &
  EM neural membrane &
  U-net & 
%  Segmentation of the EM membrane. &
  No &
  Documentation (installation), tutorials (Jupyter notebook). &\url{https://github.com/ilastik/nature\_methods\_multicut\_pipeline} \\
nucleAIzer~\cite{hollandi2020nucleaizer} &
  Fluorescent and H\&E stained nuclei &
  Mask R-CNN+U-net &
%  Nuclei segmentation. &
  No &
  Documentation (installation and application), tutorials (shell scripts). &
  \url{https://github.com/spreka/biomagdsb} \\
DenoiSeg~\cite{buchholz2021denoiseg} &
  Fluorescent cell membrane and nuclei &
  U-net &
%  Data denoise and cell membrane and nuclei segmentation. &
  No &
  Documentation (installation), tutorials (Jupyter notebook). &
  \url{https://github.com/juglab/DenoiSeg} \\
InstantDL~\cite{waibel2021instantdl} &
  Fluorescent, H\&E stained and label-free nuclei &
  U-net/Mask R-CNN &
%  Fluorescent and histology nuclei segmentation. &
  No &
  Documentation (installation and applications), dockerised. &
  \url{https://github.com/marrlab/InstantDL} \\
DeepCell~\cite{greenwald2022whole,van2016deep} &
  Fluorescent nuclei and cell membrane &
  Resnet50 &
%  Fluorescent nuclei and cell segmentation. &
  Yes &
  Documentation (application), tutorials (script), dockerised. &
  \url{https://github.com/vanvalenlab/deepcell-applications} \\
SplineDist~\cite{mandal2021splinedist} & 
  Fluorescent and H\&E stained nuclei &
  StarDist &
%  Built based on StarDist for fluorescent and histology nuclei segmentation. SplineDist is more generalised and can model non-star-convex objects with the possibility of conducting statistical shape analysis within the tool. &
  No &
  Tutorials (Jupyter notebook). &
  \url{https://github.com/uhlmanngroup/splinedist} \\
CDeep3M~\cite{haberl2018cdeep3m} & 
  XRM, ET, fluorescent and SBEM nuclei; SBEM synaptic vesicles, mitochondria and membranes &
  DeepEM3D-Net (dense CNN) &
%  CDeep3M provides a plug-and-play cloud-based deep learning solution for image segmentation of light, electron and X-ray microscopy. &
  No &
  Documentation (installation), dockerised, implemented through Amazon Web Services. &
  \url{https://github.com/CRBS/cdeep3m} \\
CellSeg~\cite{lee2022cellseg} & 
  Fluorescent nuclei and cell membrane &
  Mask R-CNN &
%  CellSeg is built for fluorescent nuclei and cell segmentation. &
  No &
  Tutorials (Jupyter notebook). &
  \url{https://github.com/michaellee1/CellSeg} \\
EmbedSeg~\cite{lalit2022embedseg} & 
  Fluorescent cell membrane and nuclei &
  Branched ERF-Net (3D) &
%  Fluorescent nuclei and cell segmentation. &
  No &
  Documentation (installation), datasets provided for reproducibility. &
  \url{https://github.com/juglab/EmbedSeg} \\
\bottomrule
\end{tabular}
\end{sidewaystable}

DL methods not only surpass conventional techniques in the segmentation of subcellular structures in microscopy images, but also exhibit a remarkable generalisation capacity, accommodating diverse imaging conditions, fluorescent markers or proteins, and cell types~\cite{caicedo2019evaluation,fischer2020mitosegnet,stringer2021cellpose,dang2023deep,hollandi2022nucleus}. This has led to the creation of several freely available tools providing pre-trained models for biologists to segment and subsequently analyse microscopy dataset quantitatively (Table~\ref{tab:subcellular}). 

% Single-cell segmentation is another main use case of deep learning, and work from our own laboratory has demonstrated that cell instances can be segmented with deep learning even with limited training data. Fig~\ref{fig:nuclei_instance_segmentation_apeer} shows an example of the nuclear instance segmentation using the built-in deep learning annotation and segmentation tool of the ZEISS arivis Cloud (formerly APEER)~\cite{apeer2023,david2021apeer} trained with only 5 cellular movies.

% \begin{figure}[!htbp]
% \centering
% \includegraphics[width=13.5cm]{fig/nuclei_instance_segmentation_apeer.pdf}
% \caption{An example of deep learning-based instance segmentation through the ZEISS arivis Cloud (formerly APEER). (A) The original cellular image for inference. (B) The instance segmentation inference result. The segmentation model was trained with only 5 cellular movies.}
% \label{fig:nuclei_instance_segmentation_apeer}
% \end{figure}

\subsection{Classification}
Classification refers to assigning text labels to images and is frequently used in cell biology and digital pathology. Instance classification focuses on recognising and categorising individual objects within an image, rather than classifying the image as a whole. DL techniques are applied to identify and classify individual cells and nuclei and to provide quantitative information on cell populations and their distribution~\cite{graham2019hover,gamper2019pannuke}. Cell type and subcellular structure identification are other applications of instance classification, which has allowed robust quantitative studies of cell function~\cite{simm2018repurposing}, cell interaction~\cite{nitta2018intelligent}, phenotype ('yes' or 'no' prediction) ~\cite{godinez2017multi} and spatial patterns and protein localisation in fluorescence images~\cite{sullivan2018deep,kraus2016classifying,kraus2017automated}. Classification has also been employed for large-scale phenotypic profiling of small molecules by analysing cellular responses to drug treatments at the single-cell level~\cite{scheeder2018machine} to evaluate drug efficacy, mechanism of action, and potential side effects. 

Manual annotations by cell biology experts are robust but time-consuming and expensive. To offset this cost, active learning~\cite{monarch2021human} has been proposed. Active learning is a powerful human-in-the-loop process in deep learning. It involves annotating manually a subset of (not all) relevant objects in images, training with this subset, and generating initial segmentation and classification masks for all instances including unannotated ones~\cite{van2021biological}. Then, the auto-generated initial segmentation and classification can be reviewed and manually corrected, serving as annotations of the next training iteration, making the human-in-the-loop process a cost-efficient approach~\cite{stringer2021cellpose,pachitariu2022cellpose,dang2023deep}.
 
 Unlike deep learning methods used for image classification, traditional machine learning (ML) based classifiers are humanly interpretable, which is important for failure analysis and model improvement~\cite{wang2021annotation}. While the DL framework has higher recognition accuracy on large sample data sets, the traditional ML approach (eg., Support Vector Machine, SVM) is thought to be a better solution for small data sets~\cite{narotamo2021machine,wang2021comparative}. So hybrid approaches that combine ML and DL techniques are being used for high accuracy and precision for cell-type classification problems~\cite{wahid2019performance,rani2022machine}, as a step towards explainable AI. 

% During the human-in-the-loop annotation process, the testing set is always unseen to the model. Fig~\ref{fig:nuclei_instance_segmentation_classification_apeer} shows a use case of nuclei instance classification through arivis AI toolkit (the built-in annotating tool) of the ZEISS arivis Cloud (formerly APEER)~\cite{apeer2023,david2021apeer} platform where we take advantage of active learning for ground truth generation.

% \begin{figure}[!htbp]
% \centering
% \includegraphics[width=9cm]{fig/nuclei_instance_segmentation_classification_apeer_4class_scale_bar_updated.pdf}
% \caption{Image segmentation using ZEISS arivis Cloud (formerly APEER) to classify nuclei atypia. (A) A raw image of RPE1 p21-GFP mRuby-PCNA cells with different nuclei atypia~\cite{hart2021multinucleation}. (B) Nuclei annotation of the image: the fluorescently labelled nuclei atypia were annotated at different time points and z-stacks using the built-in annotating tool of ZEISS arivis Cloud (formerly APEER)~\cite{apeer2023,david2021apeer}. Work from our own laboratory has demonstrated that cell instances can be segmented and classified with deep learning even with limited training data. The model trained with only 5 time-lapse movies (121 images) could successfully segment and classify normal nuclei (shown in C, pseudocoloured in blue in B), misshaped nuclei (shown in D, pseudocoloured in green in B), binucleated nuclei (shown in E, pseudocoloured in red in B), and multinucleated nuclei (shown in F, pseudocoloured in orange in B).}
% \label{fig:nuclei_instance_segmentation_classification_apeer}
% \end{figure}
\subsection{Tracking}
Tracking is the process of identifying and linking the movement of specific objects over time in a series of time-lapse images or a movie. Tracking methods in cell biology are primarily DL-independent, unlike real-world scenarios such as autonomous driving where DL-based tracking is being widely used~\cite{chen2017online,ciaparrone2020deep,marvasti2021deep,jiao2021deep,pal2021deep}. From a computational perspective, the task of tracking consists of detection-based tracking (DBT), and detection-free tracking (DFT) ~\cite{luo2021multiple}. DBT, also commonly referred to as tracking-by-detection, usually consists of two main steps: detection of the objects of interest and linking their positions and properties across consecutive frames. On the other hand, DFT requires the manual initialisation of a fixed number of objects in the first frame and then localising these objects in the subsequent frames. DBT is widely used compared to DFT since objects can be newly discovered or transiently lost through time in most scenarios and DFT cannot deal with the case~\cite{luo2021multiple}. 

In many tracking studies, deep learning is used in the detection step, including techniques such as the R-CNN series~\cite{he2017mask,girshick2015fast,ren2015faster}, YOLO~\cite{redmon2016you,redmon2018yolov3,jiang2022review} and SSD~\cite{liu2016ssd}. Deep learning can also be used for trajectory or motion prediction to support tracking. Most of the DL-based trajectory predictions are through LSTM technique~\cite{sherstinsky2020fundamentals} which has extensively progressed by predicting the coordinators of selected objects in the upcoming time frame~\cite{chandra2019traphic,chandra2019robusttp,leon2021review,wang2019exploring}. Some studies have taken advantage of convolutional feature extraction~\cite{chen2017online,he2017cell} for the prediction of trajectory. Currently~\cite{ciaparrone2020deep,marvasti2021deep,jiao2021deep,pal2021deep} the top application scenarios of DL-based tracking are pedestrian detection and autonomous vehicles, augmented reality (AR) and virtual reality (VR), and these could be brought to cell biology to advance multiscale system studies where subcellular, cellular and tissue levels changes are simultaneously modulated and measured~\cite{floerchinger2021optimizing,venkatesan2021virtual,razavian2019augmented}.

Typical examples in cell biology applications include single-cell tracking~\cite{blockhuys2020single}, multi-cell tracking during collective cell migration ~\cite{song2023machine} or particle or organelle tracking within cells~\cite{jaqaman2008robust,tinevez2017trackmate}. Tracking is challenging from both computational and biological perspectives for many reasons. First, objects can move from area to area, so each instance should be identified on a single-frame basis and these detections should be linked over time to avoid misconnections. Second, objects that are to be tracked can merge (mitochondria) or split (cell division), and this presents a discontinuity challenge in their morphology, leading to misrecognition. Third, there is a limitation in terms of the frame rate in time-lapse movies~\cite{nicovich2014acquisition,danuser2014reply}, and this makes tracking in general and in 3-dimensions, in particular, challenging due to time discontinuity. Misconnection and misrecognition challenges could be at least in part overcome using DL methods for trajectory prediction. 

Tracking subcellular structures and their changes through 3-Dimensional space is a challenging but rewarding application, as it can provide valuable insight into cell dynamics ~\cite{zulkipli2018spindle,pennycook2021palbociclib} and support systems-level modelling efforts to explore complex signalling and regulatory pathways~\cite{corrigan2015modeling,min2019spontaneously}. For example, analysing the patterns of cell movements following distinct molecular perturbations has helped dissect molecular principles that govern cellular migration~\cite{van2016deep,tsai2019usiigaci,he2017cell,mavska2023cell}. Whole-cell tracking to monitor cell or nuclear size changes and the timing and duration of cell cycle phases~\cite{dang2023deep} or intracellular tracking to analyse the movement of intracellular organelles, vesicles, or proteins, within a cell,~\cite{newby2018convolutional,tinevez2017trackmate,ritter2021deep,spilger2021deep,ritter2021deep}, have taken advantage of apriori knowledge of distinct features (structural or dynamic) which have been uniquely used to solve each individual tracking problem. 

\section{The challenges and opportunities}
\subsection{Challenges of AI-guided methods in cell dynamics studies}
\subsubsection{Lack of well-annotated datasets}
Deep learning-based approaches require large amounts of labelled (annotated) data. Ideally, high-quality cell biology data need to be annotated by experts which is time-consuming. Although crowdsourcing can offer cost-effective solutions, annotation inconsistencies can require correction by experts~\cite{spiers2021deep}. Furthermore, variations in subcellular morphologies, staining protocols, and imaging quality can make the annotation challenging for non-experts. Many solutions are being developed to tackle this challenge, including active learning~\cite{vununu2021classification,moen2019deep}, transfer learning~\cite{vununu2021classification,moen2019deep,kensert2019transfer,minoofam2021trcla} and data augmentation techniques~\cite{moen2019deep,naghizadeh2021semantic,majurski2019cell,yu2021generative,chlap2021review}. Augmentation strategies where an image is altered in scale or intensity provide additional samples without necessarily increasing the number of manually annotated samples~\cite{dang2023deep}. Karabag et al investigate the impact of the amount of training data and shape variability on the U-net-based segmentation~\cite{karabaug2023impact}. They suggest that data augmentation methodologies may not improve training if the acquired cell pairs are not representative of other cells. Therefore, a thorough investigation of various augmentations is recommended. Despite mentioned solutions, the shortage of high-quality labelled data remains a critical limitation for AI-guided images and time-lapse movie analysis. Only a limited number of open-sourced datasets are available, as listed chronologically in Table~\ref{tab:existing_dataset}.

{
\footnotesize 
\begin{table}[!htbp]
\centering
\caption{\label{tab:existing_dataset}Open-sourced datasets for cell biology image and movie analysis tasks}
\begin{tabular}{p{4cm}p{6cm}p{3cm}}
\toprule
\bfseries{Dataset} & \bfseries{Description} & \bfseries{Source} \\
\midrule
Broad Bioimage Benchmark Collection~\cite{ljosa2012annotated} &
  Over 11 million images from 52 datasets for segmentation, phenotype classification, and image-based profiling tasks. &
  \url{https://bbbc.broadinstitute.org/image\_sets} \\
ISBI cell tracking challenge dataset collection~\cite{mavska2014benchmark} &
  Ten 2D image datasets and ten 3D time-lapse movie datasets of fluorescent counterstained nuclei or cells for segmentation and tracking tasks. &
  \url{http://www.celltrackingchallenge.net/} \\
DeepCell dataset~\cite{van2016deep} &
   $\sim$75,000 single-cell annotations including live-cell movies of fluorescent nuclei ($\sim$10,000 single-cell movie trajectories over 30 frames) and static images of whole cells for segmentation tasks. &
  \url{https://github.com/vanvalenlab/deepcell-tf} \\
Image data resource (IDR)~\cite{williams2017image} &
  Over 13 million images from 118 published studies. &
  \url{https://idr.openmicroscopy.org/} \\
Human Protein Atlas~\cite{thul2017subcellular} &
  Over 80,000 high-resolution confocal Immunofluorescence images showing localisation patterns of thousands of proteins for a variety of human cell lines for segmentation tasks. &
  \url{http://www.proteinatlas.org} \\
The Cell Image Library~\cite{bray2017dataset} &
  919,874 five-channel fields of morphologies of U20S cells and populations representing 30,616 tested compounds. &
  \url{https://github.com/gigascience/paper-bray2017} \\
Salmonella-infected HeLa cells~\cite{antoniou2019high} &
   93,300 multi-channel confocal fluorescence images. &
  \url{https://dataverse.harvard.edu/dataset.xhtml?persistentId=doi:10.7910/DVN/FYGHFO} \\
JUMP cell painting datasets~\cite{chandrasekaran2022three} &
  Images of osteosarcoma cells perturbed with CRISPR-mediated knockdowns and overexpression reagents and $\sim$120,000 compounds. &
  \url{https://jump-cellpainting.broadinstitute.org/} \\
NYSCF automated deep phenotyping dataset (ADPD)~\cite{schiff2022integrating} &
  Cell painting dataset of 1.2 million images (48 TB). &
  \url{https://nyscf.org/open-source/nyscf-adpd/} \\
Poisson-Gaussian Fluorescence Microscopy Denoising Dataset~\cite{zhang2019poisson} &
  Over 12,000 fluorescence microscopy images using confocal, two-photon and widefield microscopes. & \url{https://drive.google.com/drive/folders/1aygMzSDdoq63IqSk-ly8cMq0_owup8UM} \\ 
\bottomrule
\end{tabular}
\end{table}
}

\subsubsection{The quality of image datasets}
DL models rely on extracting patterns from the dataset, making the quality of annotated data crucial. Inconsistent ground truth yields incorrect analytical results, while biased data (highlighting some but not all phenotypes) leads to incorrect patterns or inaccurate predictions. Noise intrinsic to microscopy can also increase the complexity of the model required to accurately capture the underlying features. This may lead to overfitting, where the model becomes too complex and fails to generalise to new and unseen data. Noisy data can also cause adversarial attacks on deep learning models, leading to misclassification of cell types, incorrect tracking of cell movements, and under- or over- segmentation of cells~\cite{hirano2021universal,wang2019security}. Meiniel et al. present a comprehensive review of microscopy data quality control and denoising using computational techniques~\cite{meiniel2018denoising}. To manage the problem of high-quality image availability, the image data resource has been set up to allow easy image data access, storage and dissemination~\cite{williams2017image}. Overall, it is essential to ensure that datasets used for deep learning are of high quality, with solid ground truth, minimal noise and free from bias~\cite{cimini2023twenty}.

\subsubsection{Model interpretability}
The challenge of interpretability for deep learning models arises from the complex and black-box nature of these models~\cite{castelvecchi2016can,ekanayake2022novel}. DL models can automatically extract complex features and patterns from large amounts of data through multiple layers of neurons~\cite{lecun2015deep}. While this makes such models powerful, in tasks such as image segmentation or classification, it also presents a challenge in understanding how the models arrived at their predictions or decisions. One approach to addressing this challenge is to visualise and examine the activations of individual neurons or groups of neurons within the model~\cite{montavon2018methods}. This technique provides insights into the patterns that the model has used to form its decision. However, these visualisations may be difficult to interpret without a deep understanding of the model architecture and data domain (see review~\cite{mohamed2022review} for more information). 

\subsubsection{High cost in real-world scenarios}
Deep learning-based methods are often expensive due to two main factors. First, effective training of DL models requires a large amount of data which can be expensive to generate. Second, the training process can be computationally intensive requiring high-performance computing resources, such as hardware of graphics processing units (GPUs) and tensor processing units (TPUs). This infrastructure cost can dissuade the planning of imaging studies needed to build the DL model~\cite{nogare2023bioimage}. DL model-building efforts supported by agencies/consortiums beyond individual researchers can help meet upfront costs, and maintain standards to make sure models are reusable~\cite{munappy2022data}. 

\subsubsection{The generalisability issue}
Generalisability denotes the extent to which a DL model trained on a specific dataset might perform well on new data, especially when the new data has different features or patterns compared to the training data. To showcase generalisability, DL models are deployed on data acquired from a different cell type or microscope~\cite{graham2019hover,gamper2019pannuke}. Efforts to reuse or generalise workflow are ongoing~\cite{mavska2023cell,stringer2021cellpose}. Generalisability issues arising due to sample variability or differences in image acquisition are being addressed through data augmentation, multi-task learning, swarm learning or collaboration with domain specialists~\cite{ali2022assessing,wang2019pathology,wang2019deep}. 

\subsection{Opportunities of AI-guided methods in cell dynamics studies}
With the advent of new AI-guided methods to identify, track and analyse objects in time-lapse movie datasets (Table~\ref{tab:existing_dataset}), we expect new opportunities for large-scale cell biology applications in drug discovery, drug repositioning and phenome-genome interaction map-building efforts. 

\subsubsection{Drug discovery and repositioning}
AI approaches in microscopy-based drug development or drug target identification primarily use still image datasets which are snapshots of dynamic processes ~\cite{krentzel2023deep,tran2023artificial,karacosta2021imaging}. Such still image-based drug screening efforts do not yet fully benefit from cellular and subcellular dynamics that can be visualised using high-speed live-imaging microscopes~\cite{wagner2021deep,efstathiou2021electrically,barazia2022imaging,yamashita2020digital}. Incorporating dynamic changes through time can address challenges posed by cellular heterogeneity, cell cycle stages, cell fate dissimilarities, variations in protein expression variations, cellular or subcellular dimensions, inter/intra-cellular signalling~\cite{padovani2022segmentation}. In addition to taking advantage of cell dynamics principles, AI-guided methods for movie datasets can accelerate several steps of drug discovery including cell toxicity assays~\cite{pulfer2022ades}, cell cycle profiling and morphology analysis~\cite{padovani2022segmentation,ren2021deep}. Increasing single-cell movie datasets along with the development of DL model standards can integrate image-omics with other -omics datasets that capture dynamic information and have accelerated drug repositioning studies~\cite{iorio2015semi,mertens2023drug}. Investing in collaborative efforts to compile microscopy datasets can fuel the development of robust AI-guided methods. This, in turn, will unlock research and engineering opportunities, facilitating a cyclical learning process to uncover unexplored cellular transition states in frontier biology and drug discovery studies.

\subsubsection{Genome-phenome mapping}
Genetic interaction maps built using cell biology approaches are transforming our understanding of several biological processes~\cite{chessel2019observing}, but their influence is limited to the specific model system or experimental set-up. We are just beginning to reliably link datasets from different cell types, fluorescent markers or imaging systems~\cite{roberts2017systematic,allenCell2017}. AI-guided image analysis methods are well positioned to extract information across image and video datasets, across different databases, in an unbiased form as they can be trained to search for patterns (for example, nuclear atypia such as multinucleated, misshapen and binucleated structures~\cite{hart2021multinucleation}) could be gathered across 100s of cell lines or drug treatments). Currently, high-throughput genome-phenome mapping image datasets of various cell types and models are deposited in a disconnected fashion because there is not much incentive to unify them. AI-guided methods may offer the possibility and the value in developing universal standards for collating data, in addition to existing global efforts to name and store large movie datasets~\cite{linkert2010metadata,moore2021ome,moore2023ome}.

\subsubsection{Precision medicine}
Genetic variant interpretation and classification using high-throughput cell biology methods is still a nascent field. Germline variant guidelines are well established~\cite{richards2015standards} and somatic variant guidelines are being established~\cite{parikh2020identification}. In both cases we expect single-cell imaging, the associated image dataset, and image analysis methods to play a crucial role in stratifying variant pathogenicity. To build stratification methods that are scalable, generalisable and interrogatable (crosscheck), DL models could be trained to detect and classify phenotype changes and hidden patterns. Swarm-learning has been proposed for decentralised and confidential X-ray image analysis~\cite{warnat2021swarm} and digital pathology~\cite{saldanha2022swarm} which could be extended to cell biology images and live-cell movies. As AI methods become incorporated within the clinical prognosis framework~\cite{oren2020artificial,cui2021artificial}, we predict there will be a growing demand for robust models for the clinical actionability of molecular targets in cancer therapies, genetic rare diseases and infectious diseases.

\section{Conclusion}
The impact of deep learning methods in large-scale and complex microscopy data analysis has been significant. Deep learning techniques have already revolutionised still image analysis and are now beginning to transform time-lapse movie analysis through state-of-the-art performance in a wide range of applications, such as object detection and tracking, segmentation, unsupervised clustering and classification. Deep learning methods employed to segment and classify cells are beginning to detect novel anomalies in 3D structures~\cite{balkenhol2019deep} or time series data~\cite{ji2021novel}, identify distinctive transient cellular transitions~\cite{ren2021deep}, and reveal complex behaviours and movement patterns~\cite{molina2022acme,dang2023deep} which were previously unrecognised. 

Automated and data-driven workflows and cloud-based large-scale solutions have significantly improved the speed, efficiency and accuracy of DL-guided image analysis tasks, while also increasing the ease with which biologists can implement AI tools. Overall, the use of deep learning methods in microscopy has enabled researchers to extract valuable information, some that are not obvious to the eye, from huge volumes of image data and opened up new opportunities in medical diagnosis and clinical translation. 

It is important to recognise that deep learning methods rely on abundant, robustly-annotated data and careful parameter-tuning. Assessing their reliability and interpretability can be challenging~\cite{cimini2023twenty}, which can restrict their applications in some domains. The establishment of universally accepted standards and frameworks to store and share human-annotated image datasets, DL models, and post-processing pipelines are complex challenges~\cite{nogare2023bioimage} that necessitate attention through international collaborations and consortia. 

\section*{Acknowledgement}
We thank Janeth Catalina Manjarrez-Gonzalez (Draviam group) for unpublished training images; Dr Sreenivasan Bhattiprolu (Zeiss, USA), Dr Peter Thorpe (QMUL) and Dr Nikola Ojkic (QMUL) for comments on the manuscript content and Draviam group members for useful discussions. We would like to acknowledge funding support from the Biotechnology and Biological Sciences Research Council (BBSRC) to V.M. Draviam (BB/R01003X/1, BB/T017716/1, BB/W002698/1 and BB/X511067/1), B. Chai (BB/X511067/1 and InnovateUK SBCY1K9R), and C. Efstathiou (LIDo-iCASE studentship [BB/T008709/1]). B Chai is a Knowledge Transfer Partnership associate collaborating with Zeiss UK.

\section*{Author contributions}
BC drafted the manuscript together with VMD and CE and edited sections using comments from HY. BC contributed to Fig~\ref{fig:new_fig} and Tables~\ref{tab:dl_tech}, ~\ref{tab:subcellular} and~\ref{tab:existing_dataset}; HY introduced SAM to Fig~\ref{fig:new_fig} and Table~\ref{tab:dl_tech}; CE generated Table~\ref{tab:subcellular} with support from BC and VMD.

%% file: glossary.tex
\section*{Glossary}
% According to TiCB guidance, the glossary should be in alphabetical order 

\subsection*{Data annotation}
the process of adding attributes to training data and labelling them so that a DL model can learn what predictions it is expected to make.

\subsection*{Edge-based segmentation}
a conventional segmentation approach that aims to first detect the contours of the specific object and then fill in the contours for segmentation.

\subsection*{Instance classification}
usually consists of object detection, localising their position within the image and classifying them into predefined categories.

\subsection*{Long Short-Term Memory (LSTM)}
a type of Recurrent Neural Network (RNN) architecture that was designed to overcome the problem of vanishing and exploding gradients faced by standard RNNs. LSTM is suited to tasks involving sequences with long-term dependencies, such as time series prediction, natural language processing, and speech recognition.

\subsection*{Neural network} 
a densely interconnected group of nodes. Each node connects to several nodes in the layer beneath it, from which it receives data (eg., training data in the last layer), and several nodes in the layer above it, to which it outputs data. Incoming connections are assigned weights. Active nodes multiply their respective weights and pass them forward if it exceeds a threshold. Training involves adjusting weights and thresholds are adjusted to produce similar outputs for data with the same labels.

\subsection*{Segmentation}
the process of dividing an image into multiple regions or segments with each corresponding to a specific object or area of interest.

\subsection*{Single Shot Detector (SSD)}
an object detection method that uses a single neural network for the entire image, similar to YOLO. Directly based on different image regions, it predicts bounding boxes and class probabilities directly. Unlike YOLO, SSD operates on multiple feature maps with different resolutions to handle objects of various sizes.

\subsection*{Thresholding segmentation}
a conventional segmentation method by choosing a threshold based on the intensity histogram for segmenting an object.

\subsection*{You Only Look Once (YOLO)}
an object detection method with the key idea of applying a single neural network to the full image, which divides the image into regions and predicts bounding boxes and probabilities for each region.

\subsection*{Zero-shot learning}
a remarkable ML/DL method which refers to recognising new unseen objects, and so it can be applied to new image distributions and tasks.